\documentstyle[epsfig]{mn}

\newif\ifAMStwofonts

\ifoldfss
  \ifCUPmtlplainloaded \else
    \NewTextAlphabet{textbfit} {cmbxti10} {}
    \NewTextAlphabet{textbfss} {cmssbx10} {}
    \NewMathAlphabet{mathbfit} {cmbxti10} {} 
    \NewMathAlphabet{mathbfss} {cmssbx10} {} 
  \fi
  \ifAMStwofonts
    \ifCUPmtlplainloaded \else
      \NewSymbolFont{upmath} {eurm10}
      \NewSymbolFont{AMSa} {msam10}
      \NewMathSymbol{\upi}     {0}{upmath}{19}
      \NewMathSymbol{\umu}     {0}{upmath}{16}
      \NewMathSymbol{\upartial}{0}{upmath}{40}
      \NewMathSymbol{\leqslant}{3}{AMSa}{36}
      \NewMathSymbol{\geqslant}{3}{AMSa}{3E}

      \let\leq=\leqslant 
      \let\geq=\geqslant 
    \fi
  \fi
\fi 

\ifnfssone
  \newmathalphabet{\mathit}
  \addtoversion{normal}{\mathit}{cmr}{m}{it}
  \addtoversion{bold}{\mathit}{cmr}{bx}{it}
  \newmathalphabet{\mathbfit} 
  \addtoversion{normal}{\mathbfit}{cmr}{bx}{it}
  \addtoversion{bold}{\mathbfit}{cmr}{bx}{it}
  \newmathalphabet{\mathbfss} 
  \addtoversion{normal}{\mathbfss}{cmss}{bx}{n}
  \addtoversion{bold}{\mathbfss}{cmss}{bx}{n}
  \ifAMStwofonts
    \ifCUPmtlplainloaded \else
      \UseAMStwoboldmath
      \makeatletter
      \new@mathgroup\upmath@group
      \define@mathgroup\mv@normal\upmath@group{eur}{m}{n}
      \define@mathgroup\mv@bold\upmath@group{eur}{b}{n}
      \edef\UPM{\hexnumber\upmath@group}
      \new@mathgroup\amsa@group
      \define@mathgroup\mv@normal\amsa@group{msa}{m}{n}
      \define@mathgroup\mv@bold\amsa@group{msa}{m}{n}
      \edef\AMSa{\hexnumber\amsa@group}
      \makeatother
      \mathchardef\upi="0\UPM19
      \mathchardef\umu="0\UPM16
      \mathchardef\upartial="0\UPM40
      \mathchardef\leqslant="3\AMSa36
      \mathchardef\geqslant="3\AMSa3E

      \let\leq=\leqslant 
      \let\geq=\geqslant 
    \fi
  \fi
\fi 

\ifnfsstwo
  \DeclareMathAlphabet{\mathbfit}{OT1}{cmr}{bx}{it}
  \SetMathAlphabet\mathbfit{bold}{OT1}{cmr}{bx}{it}
  \DeclareMathAlphabet{\mathbfss}{OT1}{cmss}{bx}{n}
  \SetMathAlphabet\mathbfss{bold}{OT1}{cmss}{bx}{n}
  \ifAMStwofonts
    \ifCUPmtlplainloaded \else
      \DeclareSymbolFont{UPM}{U}{eur}{m}{n}
      \SetSymbolFont{UPM}{bold}{U}{eur}{b}{n}
      \DeclareSymbolFont{AMSa}{U}{msa}{m}{n}
      \DeclareMathSymbol{\upi}{0}{UPM}{"19}
      \DeclareMathSymbol{\umu}{0}{UPM}{"16}
      \DeclareMathSymbol{\upartial}{0}{UPM}{"40}
      \DeclareMathSymbol{\leqslant}{3}{AMSa}{"36}
      \DeclareMathSymbol{\geqslant}{3}{AMSa}{"3E}

      \let\leq=\leqslant 
      \let\geq=\geqslant 
    \fi
  \fi
\fi 

\ifCUPmtlplainloaded \else
  \ifAMStwofonts \else 
    \def\upi{\pi}
    \def\umu{\mu}
    \def\upartial{\partial}
  \fi
\fi

\title[Sub-mm observations of HLIRGs]
  {Sub-millimetre observations of hyperluminous infrared galaxies}
\author[D. Farrah et al.]
  {D.~Farrah$^1$, S.~Serjeant$^{2}$, A.~Efstathiou$^3$, M.~Rowan-Robinson$^1$, A.~Verma$^4$\\
 $^1$Astrophysics Group, Blackett Laboratory, Imperial College, Prince Consort Road, London SW7 2BW, UK\\
 $^2$Unit for Space Sciences \& Astrophysics, School of Physical Sciences, University of Kent at Canterbury,\\
     Canterbury, Kent, CT2 7NZ, UK\\
 $^3$Department of Computer Science and Engineering, Cyprus College, 6 Diogenous Street, PO Box 22006, \\
     1516 Nicosia, Cyprus\\
 $^4$Max-Planck-Institut fur Extraterrestrische Physik, Postfach 1312, 85741 Garching, Germany}
\date{Received 2001 October 31}
\pagerange{\pageref{01}--\pageref{16}}
\pubyear{2001}

\def\LaTeX{L\kern-.36em\raise.3ex\hbox{a}\kern-.15em
    T\kern-.1667em\lower.7ex\hbox{E}\kern-.125emX}

\begin{document}

\label{firstpage}

\maketitle

\begin{abstract}

We present sub-mm photometry for 11 Hyperluminous Infrared Galaxies (HLIRGs, $L_{IR} > 10^{13.0} h_{65}^{-2} L_{\sun}$) 
and use radiative transfer models for starbursts and AGN to examine the nature of the IR emission. In all the sources both a 
starburst and AGN are required to explain the total IR emission. The mean starburst fraction is $35\%$, with a range spanning $80\%$ 
starburst dominated to $80\%$ AGN dominated. In all cases the starburst dominates at rest-frame wavelengths longwards of $50\mu$m, 
with star formation rates $>500M_{\odot}$yr$^{-1}$. The trend of increasing AGN fraction with increasing IR luminosity observed in 
IRAS galaxies is observed to peak in the HLIRG population, and not increase beyond the fraction seen in the brightest 
ULIRGs. The AGN and starburst luminosities correlate, suggesting that a common physical factor, plausibly the dust masses, 
govern the luminosities of starbursts and AGN in HLIRGs. Our results suggest that the HLIRG population is 
comprised both of ULIRG-like galaxy mergers, and of young galaxies going through their maximal star formation periods 
whilst harbouring an AGN. The detection of coeval AGN and starburst activity in our sources implies that 
starburst and AGN activity, and the peak starburst and AGN luminosities, can be coeval in active galaxies generally. 
When extrapolated to high-$z$ our sources have comparable observed frame sub-mm fluxes to sub-mm survey sources. At least 
some high-$z$ sub-mm survey sources are therefore likely to be comprised of similar galaxy populations to those found in 
the HLIRG population. It is also plausible from these results that high-$z$ sub-mm sources harbour heavily obscured AGN. 
The differences in X-ray and sub-mm properties between HLIRGs at $z\sim1$ and sub-mm sources at 
$\sim3$ implies some level of evolution between the two epochs. Either the mean AGN obscuration level is greater at 
$z\sim3$ than at $z\sim1$, or the fraction of IR-luminous sources at $z\sim3$ that contain AGN is smaller than that at 
$z\sim1$. 
\end{abstract}


\section{Introduction}
One of the most important results from the Infrared Astronomical Satellite ({\em IRAS}) all 
sky surveys was the detection of a new class of galaxy where the bulk of the bolometric emission 
lies in the infrared waveband \cite{soi,sa1}. This population, termed 'Luminous Infrared Galaxies' (LIGs), 
becomes the dominant extragalactic population at IR luminosities above $10^{11}L_{\sun}$, 
with a higher space density than all other classes of galaxy of comparable bolometric luminosity. 

At the brightest end of the LIG population 
lie the Hyperluminous Infrared Galaxies (HLIRGs), those with $L_{IR} > 10^{13.0} h_{65}^{-2} L_{\sun}$ 
\cite{rr2}. The first HLIRG to be found (P09104+4109) was identified by Kleinmann et al. \shortcite{kl}, 
with a far infrared luminosity of $1.5\times10^{13}h_{50}^{-2}L_{\sun}$. In 1991, Rowan-Robinson et al. 
identified F10214+4724 at $z=2.286$, with an apparent far infrared luminosity of $3\times10^{14}h^{-2}_{50}L_{\sun}$. 
Later observations of this object revealed a huge mass of molecular gas ($10^{11}h_{50}^{-2}M_{\sun}$ 
\cite{br,so3}), a Seyfert emission spectrum \cite{el}, high optical polarisation \cite{la}, and 
evidence for lensing with a magnification of about 10 at infrared wavelengths 
\cite{gra,bro,ei,gr2}. These objects appeared to presage an entirely new class of infrared galaxy.

The source and trigger of the IR emission in HLIRGs is currently the subject of considerable 
debate. HLIRGs may simply be the high luminosity tail of the ULIRG population, where mergers between 
evolved galaxies trigger dust shrouded starburst and AGN activity \cite{sa1}. 
There is evidence from {\it HST} observations \cite{far1} that at least some HLIRGs are merging galaxies. 
A second possibility is that HLIRGs may be very young, or 'primeval' galaxies. Rowan-Robinson 
\shortcite{rr2} argues that the majority of the emission at rest-wavelengths $>50\mu$m in 
HLIRGs is due to starburst activity, implying star formation rates $>1000M_{\sun}yr^{-1}$. If the 
rest-frame far infrared and sub-mm emission from HLIRGs is due to star formation, then the  
star formation rates would be the highest for any objects in the Universe. This would strongly suggest 
these galaxies are going through their maximal star formation periods, implying that they are galaxies 
in the first stages of formation. A final possibility is that the IR emission arises via some other 
mechanism (e.g. a transient IR luminous phase in QSO evolution not triggered by interactions), 
HLIRGs would then be an entirely different class of object.

In this paper we study the infrared emission from HLIRGs using data from the optical to the sub-mm. 
We present new sub-mm data for a sample of 11 HLIRGs, and use radiative transfer models for starbursts 
and AGN, in conjunction with previously published IR photometry, to examine the power source behind the 
IR emission. Sample selection, observations and data analysis are described in \S 2. The radiative 
transfer models used to evaluate the IR emission are described in \S 3. Results are presented in \S 4, 
and notes on individual sources are given in \S 5. Discussion is presented in \S 6 and conclusions are 
summarized in \S 7. We have taken $H_{0}=65$ km s$^{-1}$ Mpc$^{-1}$, 
$\Omega=1.0$ and $\Omega_{\Lambda}=0.0$.

\begin{table*}
\begin{minipage}{145mm}
\caption{Hyperluminous Galaxies observed by SCUBA \label{scubaobs}}
\begin{tabular}{@{}lcrlccc}
\hline
Name             & RA (J2000) & Dec                      & $z$    & Type$^{1}$ & Obs. Date    & Calibrator \\
                 & hh:mm:ss   & \degr\ \arcmin\ \arcsec\ &        &            &              &            \\
\hline
IRAS F00235+1024        & 00 26 06.7 & 10 41 27.6               & 0.58   &   nl       & October 2000 & Uranus     \\     
IRAS 07380-2342         & 07 40 09.8 & -23 49 57.9              & 0.29   &   nl       & January 2001 & oh 231     \\
IRAS F10026+4949        & 10 05 52.5 & 49 34 47.8               & 1.12   &   Sy1      & January 2001 & oh 231     \\   
IRAS F12509+3122        & 12 53 17.6 & 31 05 50.5               & 0.78   &   QSO      & January 2001 & Mars       \\
IRAS 13279+3401         & 13 30 15.3 & 33 46 28.7               & 0.36   &   QSO      & January 2001 & Mars       \\
IRAS 14026+4341         & 14 04 38.8 & 43 27 07.2               & 0.32   &   Sy1      & January 2001 & Mars       \\
IRAS F14218+3845        & 14 23 55.5 & 38 31 51.3               & 1.21   &   QSO      & January 2001 & Mars       \\
IRAS F16124+3241        & 16 14 22.1 & 32 34 03.7               & 0.71   &   nl       & January 2001 & Mars       \\
ELAISP90 J164010+410502 & 16 40 10.2 & 41 05 22.1               & 1.10   &   QSO      & January 2001 & Mars       \\   
IRAS 18216+6418         & 18 21 57.3 & 64 20 36.4               & 0.30   &   Sy1      & October 2000 & Uranus     \\
IRAS F23569-0341        & 23 59 33.6 & -03 25 12.8              & 0.59   &   nl       & October 2000 & Uranus     \\
\hline
\end{tabular}

\medskip

Positions and redshifts are taken from the NASA Extragalactic Database. 
$^{1}$Spectral Type, taken from Rowan-Robinson 2000 \& NED. 'nl' - Narrow Line object, 'Sy1' - Seyfert 1.

\end{minipage}
\end{table*}

\section{The Sample}

\subsection{Sample Selection and Observations}
The objects in our sample are taken from the FIR selected sample presented by Rowan-Robinson 
\shortcite{rr2}. Unlike nearly all previous studies of HLIRGs, the sample in our study is selected in a 
manner independent of obscuration, inclination or AGN content. Together with the statistical homogeneity 
and completeness of the parent samples, our sample is therefore entirely free from AGN bias and suitable 
for drawing global conclusions about the HLIRG population. 

Observations were made using the Submillimetre Common User Bolometer Array (SCUBA, Holland et al. 1999) on the 
James Clerk Maxwell Telescope (JCMT) on October 11-12 2000 and on January 8-17 2001. SCUBA contains two bolometer 
arrays, one containing 37 pixels and optimized for observations at $850\mu$m, and the other containing 91 pixels 
and optimized for observations at $450\mu$m. In most circumstances, both arrays are operated simultaneously using a 
dichroic. Observations were performed using SCUBA's photometry mode, in which data is taken using only a single pixel on each 
array. For each integration the secondary mirror was jiggled so that the selected bolometer in each array sampled 
a $3\times3$ grid with 2\arcsec\  spacing between grid positions, centred on the source. During each integration the secondary 
mirror is chopped by 45'' in azimuth with a frequency of 7Hz in order to remove sky variations. After each integration, 
the telescope is then nodded to a reference position 45'' away in azimuth to remove hotspots in the internal 
SCUBA optics. Each object was observed for approximately 40 minutes, depending on weather 
conditions. Sky opacities during the observations were of moderate quality, with measured opacities at 225Ghz 
from the Caltech Submillimetre Observatory (CSO) in the range $0.075 < \tau_{225} < 0.15$. Calibration observations were 
made of Mars or Uranus, or of a secondary calibrator if no primary calibrator was available. Skydips were taken before 
and after each object and calibrator observation.

\subsection{Data Reduction}
The SCUBA User Reduction Facility (SURF) software was used to reduce the data for all objects. The data were 
first flatfielded and despiked. Atmospheric extinction corrections at $450\mu$m and $850\mu$m were derived by 
extrapolating from the CSO $\tau_{225}$ extinction values following the prescription of Archibald, Wagg \& 
Jenness \shortcite{awj}. These extrapolated values were checked for consistency against the observed $450\mu$m and 
$850\mu$m extinctions from the skydips. Residual sky gradients not removed by nodding and chopping were removed by 
averaging over all the bolometers in each array, and subtracting this value from the measured source flux. Individual 
integrations for each source were then concatenated into a single exposure. Each concatenated dataset was checked 
for internal consistency using a Kolmogorov-Smirnov (K-S) test. Finally, flux calibration for each source was 
carried out using the FLUXES package together with the calibrators listed in Table \ref{scubaobs}. In those cases 
where the calibrator has a larger spatial extent than the 9 point photometry mode jigglemap the flux of the 
calibrator in a single photometry jigglemap was used to calibrate the sources, rather than the total flux of 
the calibrator.

\section{Infrared Emission Models}

\subsection{Starburst Models}

To model the IR emission due to starburst activity we used the starburst models of Efstathiou, Rowan-Robinson \& 
Siebenmorgen \shortcite{ef1}. These models consider an ensemble of evolving HII regions containing hot young stars, 
embedded within Giant Molecular Clouds (GMCs) of gas and dust. The composition of the dust is given by the dust 
grain model of Siebenmorgen \& Krugel \shortcite{sie}, which includes Polycyclic Aromatic Hydrocarbons. The stellar 
populations within the GMCs evolve according to the stellar synthesis codes of Bruzual \& Charlot \shortcite{bru}. The star 
formation rate is assumed to decay exponentially with an e-folding timescale of $2\times10^{7}$ years. The models 
vary in starburst age from zero years up to $7.2\times10^{7}$ years, with 11 discrete values.

\subsection{AGN Torus Models}

To model the IR emission due to AGN, we used the AGN models of Efstathiou \& Rowan-Robinson \shortcite{ef0}.
These models incorporate accurate solutions to the axially symmetric radiative-transfer problem in dust clouds to model 
the IR emission from dust in active galactic nuclei. Dust composition is given by the multigrain dust model of 
Rowan-Robinson \shortcite{rra}. We have used the thick tapered disk models following an $r^{-1}$ density distribution, as this 
subset of models has been found to be most successful in satisfying the observational constraints of AGN. The AGN models 
vary in torus opening angle from $0\degr$ to $90\degr$, with 15 discrete values.

\section{Results}

\subsection{Spectral Energy Distributions}
We combined our measured sub-mm fluxes and $3\sigma$ upper limits with optical and IR data from the literature to fit 
Spectral Energy Distributions (SEDs) for each object. For each source we obtained {\em IRAS} fluxes and $3\sigma$ upper 
limits from the Faint Source Catalogues \cite{mos}, or by using the SCANPI v5.0 software. Where available, we obtained 
{\em B} and {\em R} band magnitudes from the APM catalogues, and 
near-IR fluxes from the 2MASS catalogues. Additional photometry for each source is described in section 6. The compiled 
fluxes for each object are given in Table \ref{hlirgphota}.

Emission from unobscured population II and III stars is not included in either the starburst or AGN models, and care must be taken in 
accounting for this. For example, a recent {\it HST} study of ULIRGs \cite{far2} found that the optical emission was in most cases 
dominated by old stellar populations and not by light from a starburst or AGN. To avoid the uncertainties in 
assuming an arbitrary SED for an evolved stellar population, we have assumed that all emission shortwards of $4\mu$m in the 
rest frame of the objects contains a significant but unquantified contribution from old stellar populations. 
As this contribution could lie between $0\%$ and $100\%$, any measured flux at a rest-frame wavelength shorter than 
$4\mu$m is treated as a  $3\sigma$ upper limit in the fitting, unless the contribution from the host galaxy has been previously removed.

Goodness of fit was examined by using the reduced $\chi^{2}$ statistic. Fits for all sources were good, with $\chi^{2}_{best}\leq1.5$ in 
all cases. To determine the errors on the luminosities and model parameters we explored $\chi^{2}$ space between $\chi^{2}_{best}$ and 
$\chi^{2}_{best}+2$. The SEDs for those sources with sufficient IR photometry to constrain the shape of the SED are given in Figure 
\ref{hlirg_seds1}. The total IR luminosities for each source, the starburst and AGN components, and
the best fit model parameters are given in Table \ref{hlirgparams}.

There exist models for the IR emission from dusty QSOs \cite{rr95} where the dust can extend several hundred parsecs from the nucleus 
instead of the several tens of parsecs found in other AGN models \cite{ef0}. The dust at these large radii will have a temperature of 
20 - 70 Kelvin, as opposed to the dust several tens of parsecs from the AGN, which will have a temperature of a few hundred Kelvin. If  
extended dust around an AGN were present in any of our sample the sub-mm emission from this dust could be mistakenly ascribed to star 
formation, thus overestimating the star formation rates. To test this, we fitted the  models of Rowan-Robinson \shortcite{rr95} to the 
objects in our sample, both on their own and in combination with the starburst models described previously. In all cases the extended 
AGN dust models were clearly rejected by the fitting. We therefore conclude that sub-mm emission from extended dust surrounding an AGN 
is not viable for the objects in our sample.

\begin{table*} 
\footnotesize 
\centering 
\caption[]{Compiled Photometry from the optical to the sub-mm for the HLIRGs in our sample \label{hlirgphota}}    
\begin{tabular}{lcccccccccccc}\hline  
Name              & $S_{I}$& $S_{J}$ & $S_{H}$ & $S_{K}$ & $S_{6.7}$     & $S_{12}$     & $S_{25}$     & $S_{60}$     & $S_{100}$     & $S_{450}$ & $S_{850}$     \\ 
\hline
F00235+1024$^{a}$ & $0.08$ &  --     &  --     &   --    & $0.92\pm0.38$ & $<173$       & $<193$       & $428\pm55.6$ & $<938$        & $<967$    & $<6.38$       \\
07380-2342$^{b}$  & $1.11$ &  --     &  --     &  --     &   --          & $484\pm28$   & $801\pm80$   & $1170\pm88$  & $3546\pm280$  & $<1275$   & $26.8\pm4.2$  \\
F10026+4949       & $0.04$ &  --     &  --     &  --     &  --           & $<86$        & $177\pm35$   & $266\pm30$   & $310\pm80$    & $<386$    & $<7.91$       \\ 
F12509+3122$^{c}$ & $0.86$ & $0.96$  & $0.81$  & $1.13$  &     --        & $<106$       & $103\pm26$   & $218\pm44$   & $309\pm82$    & $<333$    & $<9.23$       \\
13279+3401$^{d}$  &  --    &   --    &   --    &   --    &    --         & $<94$        & $<126$       & $1182\pm77$  & $1196\pm165$  & $<274$    & $<9.22$       \\
14026+4341$^{e}$  &  --    & $3.36$  & $4.76$  & $8.19$  &    --         & $118\pm27.1$ & $285\pm29$   & $622\pm57.0$ & $994\pm239$   & $<94$     & $<7.53$       \\
F14218+3845$^{f}$ & $0.05$ &   --    &  --     &  --     & $0.79\pm0.26$ & $<96.9$      &  $<74.9$     & $<565$       & $<2100$       & $<253$    & $<8.55$       \\
F16124+3241$^{g}$ &  --    &  --     &     --  & $0.15$  &    --         & $<65$        & $<55$        & $174\pm35$   & $290\pm90$    & $<103$    & $8.47\pm2.2$  \\
E J1640+41$^{h}$  & --     & $0.44$  & $0.63$  & $0.56$  & $4.99\pm0.66$ & $<102$       & $<57$        & $<81$        & $<336$        & $<82$     & $<6.97$       \\
18216+6418$^{i}$  & --     & $8.8$   & $13.0$  & $22.1$  & $109\pm11$    & $211\pm14$   & $395\pm45$   & $1128\pm113$ & $2159\pm252$  & $<308$    & $14.8\pm2.6$  \\
\hline  
\end{tabular} 
\\
All fluxes are given in mJy. Unless stated in the text, all fluxes at rest-frame wavelengths shortward of $4\mu$m are treated as upper limits in 
the SED fitting, due to the unquantified contribution from the host galaxy. Sources for additional photometry are 
described in \S6. Additional photometry:
(a) $S_{15}=6.752\pm2.143$, $S_{90}=477.9\pm147.9$, $S_{180}=804.7\pm393.3$
(b) $S_{B}=0.062$, $S_{R}=0.346$
(c) $S_{B}=1.41$, $S_{V}=0.751$, $S_{R}=0.72$ 
(d) $S_{B}=0.50$, $S_{R}=2.38$ 
(e) $S_{B}=3.23$, $S_{R}=5.48$, $S_{4.0}=16$, $S_{7.0}=36\pm4$
(f) $S_{B}=0.06$, $S_{R}=0.114$, $S_{15}=3.228\pm1.036$, $S_{90}=163.5\pm61$, $S_{180}=<1765$
(g) $S_{B}=0.014$, $S_{R}=0.079$
(h) $S_{15}=7.75\pm0.72$, $S_{90}=72\pm23$
(i) $S_{U}=9.04$, $S_{B}=8.12$, $S_{V}=7.55$, $S_{L}=60.5$, $S_{9.63}=143\pm14.3$, $S_{1100}=<47$, $S_{1250}=<3.7$
\end{table*}

\subsection{Star Formation Rates}
Estimating obscured star formation rates from IR data is based on silicate and graphite dust grains absorbing 
the optical and UV light from young stars and reradiating in the IR and sub-mm. A recent estimate for deriving 
star formation rates from IR luminosities has been made by Rowan-Robinson et al. \shortcite{rr0}:

\begin{equation}
\stackrel{.}{M}_{*,all} = 2.6\times10^{-10}\frac{\phi}{\epsilon}\frac{L_{60}}{L_{\odot}}
\label{equnsfr}
\end{equation}

\noindent where $\stackrel{.}{M}_{*,all}$ is the rate of formation of stars, $\epsilon$ is the fraction of 
optical/UV light from the starburst that is absorbed by the dust and re-emitted in the IR, and $L_{60}$ is the 
$60\mu$m starburst luminosity. The factor $\phi$ incorporates the correction between a Salpeter IMF and the 
true IMF ($\times1.0$ for a Salpeter IMF, $\times3.3$ for a Miller-Scalo IMF), and 
also a correction for the assumed upper and lower stellar mass bounds for the stars forming in the starburst 
($\times1.0$ if the mass range is $0.1 < M_{\odot} < 100$, $\times 0.323$ if the mass range is assumed to form 
only OBA type stars, i.e.  $1.6 < M_{\odot} < 100$). We have assumed that all of the 
optical/UV light from the starburst is absorbed ($\epsilon\sim1.0$), and that the IMF of the starburst is 
a Salpeter IMF forming stars across the mass range $0.1 < M_{\odot} < 100$, ($\phi=1.0$). The calculated 
star formation rates for our sample using Equation \ref{equnsfr} are given in Table \ref{hlirgparams}.

\subsection{Dust Masses and Temperatures}
Dust temperatures cannot be directly determined from the starburst or AGN models described in sections 3.1 and 3.2, 
as these models do not assume an isothermal temperature distribution. Monolithic dust temperatures, although unphysical 
simplifications in AGN, can serve as a useful comparison with previous work. Dust temperature estimates 
were calculated by fitting an optically thin greybody function of the form:

\begin{equation}
S_{\nu} = \nu^{\beta}B(\nu, T_{dust})
\label{greybody}
\end{equation}

\noindent to the FIR SED over the wavelength range $200 - 1000\mu$m. In Equation \ref{greybody} $S_{\nu}$ is the source 
flux at a frequency $\nu$, $T_{dust}$ is the dust temperature, $B(\nu, T_{dust})$ is the Planck function, and $\beta$ 
is the frequency dependence of the grain emissivity. 

Gas and dust masses can be estimated directly from the starburst models. The total mass of Giant Molecular Clouds 
(GMC's, which include both gas and dust) is straightforward to estimate, as stars are assumed to form from the 
GMC's. The mass of stars at time $t^{*}$ can be written:

\begin{equation}
M_{s}(t^{*}) = \eta M_{GMC}(t^{*}) = \int_{0}^{t^{*}}\stackrel{.}{M}_{s}(t)dt
\end{equation}

\noindent where $M_{GMC}(t^{*})$ 
is the mass of Giant Molecular Clouds at time $t^{*}$, $\stackrel{.}{M}_{s}(t)$ is the star 
formation rate and $\eta$ is the efficiency of conversion of GMC's to stars. Approximating 
the star formation rate as an exponential decay with a characteristic e-folding time 
of $\tau=20$Myr (ERS 00), then it is trivial to show that:

\begin{equation}
M_{GMC}(t^{*}) = \frac{\tau\stackrel{.}{M}_{s}(t^{*})}{\eta}\left[\frac{1}{e^{-\frac{t^{*}}{\tau}}}-1\right]
\label{equngmc}
\end{equation}

\noindent where $M_{GMC}$ is given in multiples of $10^{6}M_{\odot}$ as $\tau$ is given in multiples of 
$10^{6}$ years. This however gives the total mass of GMCs in the system at the current age of the starburst, without 
accounting for the effects of supernovae winds, which will 
remove some fraction of gas and dust from the system either via photodestruction of dust grains or by blowing 
the gas and dust out of the starburst regions. Estimates using Equation \ref{equngmc} may therefore significantly overestimate 
the gas and dust masses in the system, unless the starburst is very young. For the purposes of this analysis 
it is better to estimate dust masses directly from the sub-mm SED. An approach for estimating dust and gas masses 
in galaxies using sub-mm data is to assume the system is optically thin at these wavelengths, and to follow the prescription 
of Hildebrand \shortcite{hil}:

\begin{equation}
M_{dust}=\frac{1}{1+z}\frac{S_{\nu_{o}}D_{L}^{2}}{\kappa(\nu_{r})B(\nu_{r},T_{dust})}
\label{eqn:mdust}
\end{equation} 

\noindent where $\nu_{o}$ and $\nu_{r}$ are the observed and rest frame frequencies respectively, $S_{\nu_{o}}$ is the 
flux in the observed frame, $B(\nu_{r},T_{dust})$ is the Planck function in the rest frame and $T_{dust}$ is the dust 
temperature. The gas mass is then obtained by assuming a fixed gas to dust ratio. For the most extreme {\it IRAS} galaxies, 
the best current estimate of the gas to dust ratio is $540\pm290$ \cite{sand}. For comparison, the gas to dust ratio 
in spiral galaxies is thought to be $\sim500$ \cite{dev}, and $\sim700$ in ellipticals \cite{wik}. The mass absorption 
coefficient in the rest frame, $\kappa_{r}$, is taken to be:

\begin{equation}
\kappa_{r}= 0.067\left(\frac{\nu_{r}}{2.5\times10^{11}}\right)^{\beta}
\label{eqn:mabsorb}
\end{equation} 

\noindent in units of m$^{2}$kg$^{-1}$ and the luminosity distance is assumed to be of the form:

\begin{equation}
D_{L}=\frac{c}{H_{0}q_{0}^{2}}(q_{0}z + (q_{0} - 1)[(2q_{0}z + 1)^{\frac{1}{2}} - 1])
\label{eqn:lumdist}
\end{equation} 

\noindent Dust masses and temperatures for our sample calculated using Equations \ref{greybody} and \ref{eqn:mdust}
are given in Table \ref{hlirgparams}.

\subsection{Individual Sources}

\subsubsection{IRAS F00235+1024}
This narrow line object is not detected in the X-ray to a limit of $L_{X}/L_{bol}=2.3\times10^{-4}$, indicating 
either atypically weak X-ray emission or an obscuring column of $N_{H}>10^{23}$cm$^{-2}$ \cite{wi}. 
{\em HST} imaging \cite{far1} shows an interacting object, with no optical QSO. {\em ISO} observations 
\cite{ver} imply that the IR emission is dominated ($>60\%$) by a starburst, but that an AGN contribution 
is required to explain the MIR emission. 

The best fit SED model for this object is given in Figure \ref{hlirg_seds1}. Additional {\em I} band data 
was taken from Farrah et al. \shortcite{far1}, and additional {\em ISO} fluxes were compiled from Verma et al. 
\shortcite{ver}. The SED of this object is best explained by a combination of 
starburst and AGN components. The {\it ISO} detections at $6.7\mu$m and $15\mu$m combined with the sub-mm upper 
limits make it impossible to fit this source with a pure starburst. We find that the 
AGN and starburst components contribute comparably to the total IR emission. The starburst component dominates at 
wavelengths $>60\mu$m. The best fit model parameters are a dust temperature of 35 Kelvin, 
a dust mass of $1.95\times10^{8}M_{\odot}$, a star formation rate of $\sim1000M_{\odot}$yr$^{-1}$, and a line of sight 
angle to the AGN torus of $<17.2^{\circ}$. The 
strong mid-IR absorption feature at $15\mu$m predicted by the 
combined SED fit results from viewing an AGN dust torus almost edge on through intervening absorption by cooler dust. 

\begin{figure*}
\begin{minipage}{170mm}
\epsfig{figure=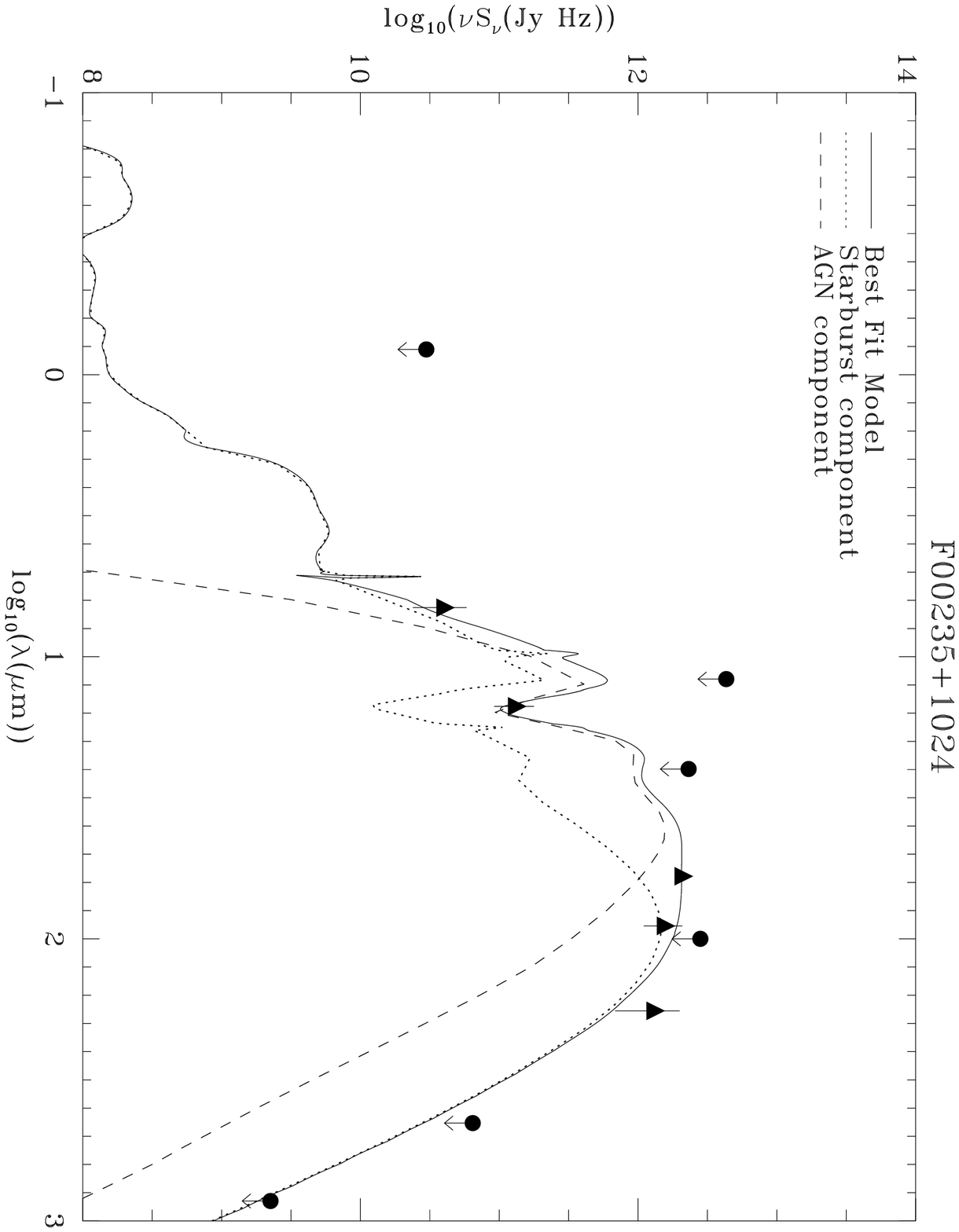,width=84mm}
\epsfig{figure=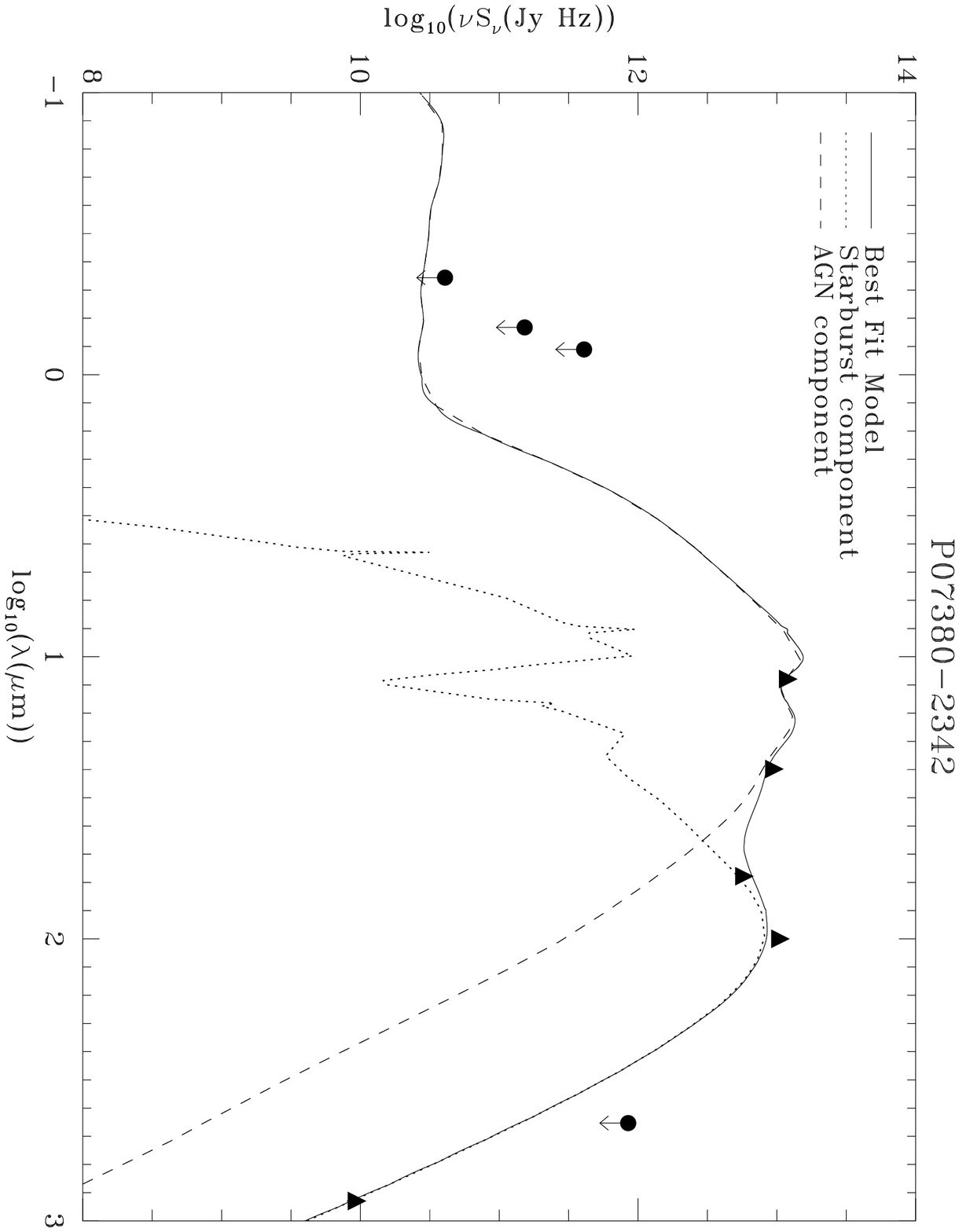,width=84mm}
\end{minipage}
\begin{minipage}{170mm}
\epsfig{figure=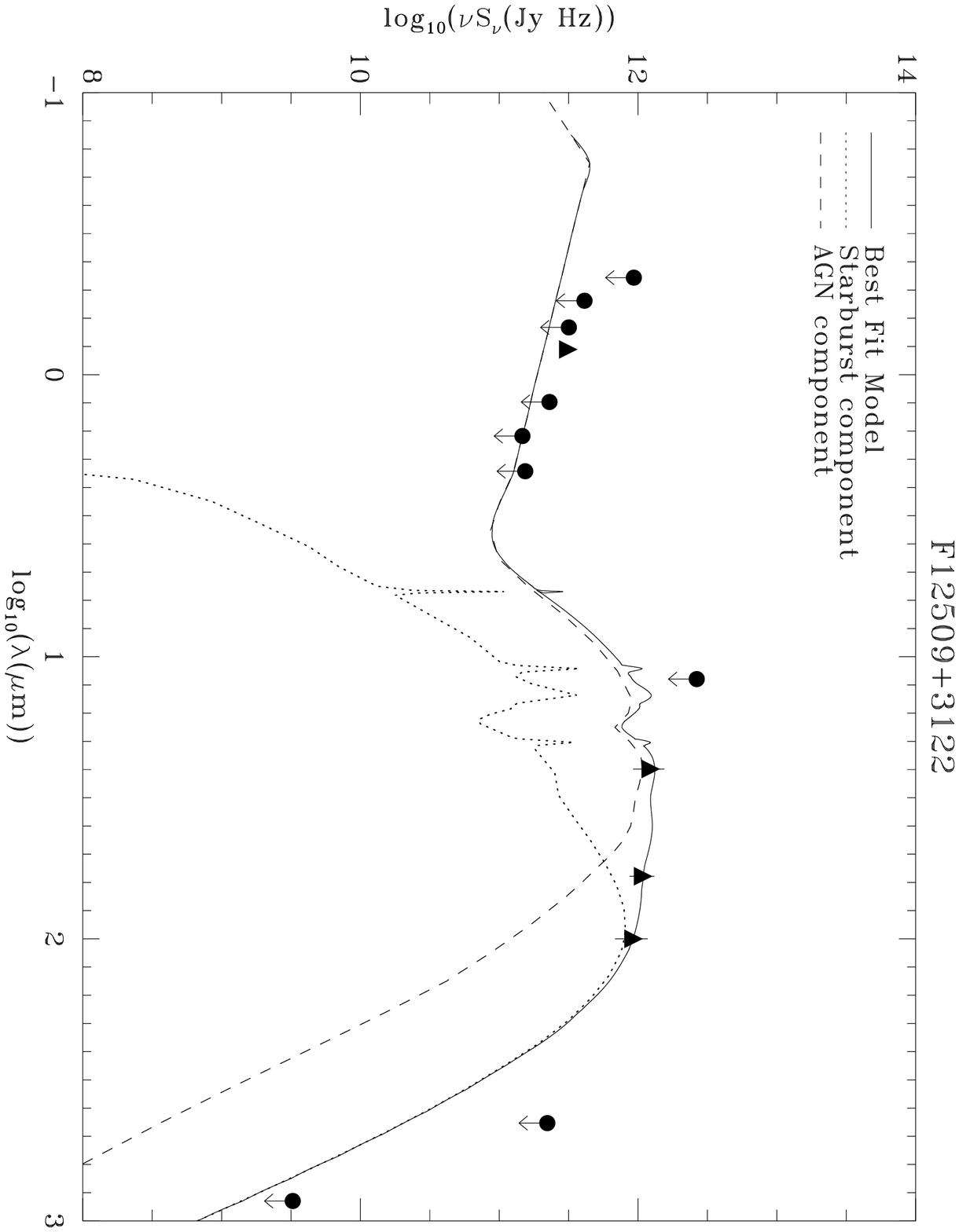,width=84mm}
\epsfig{figure=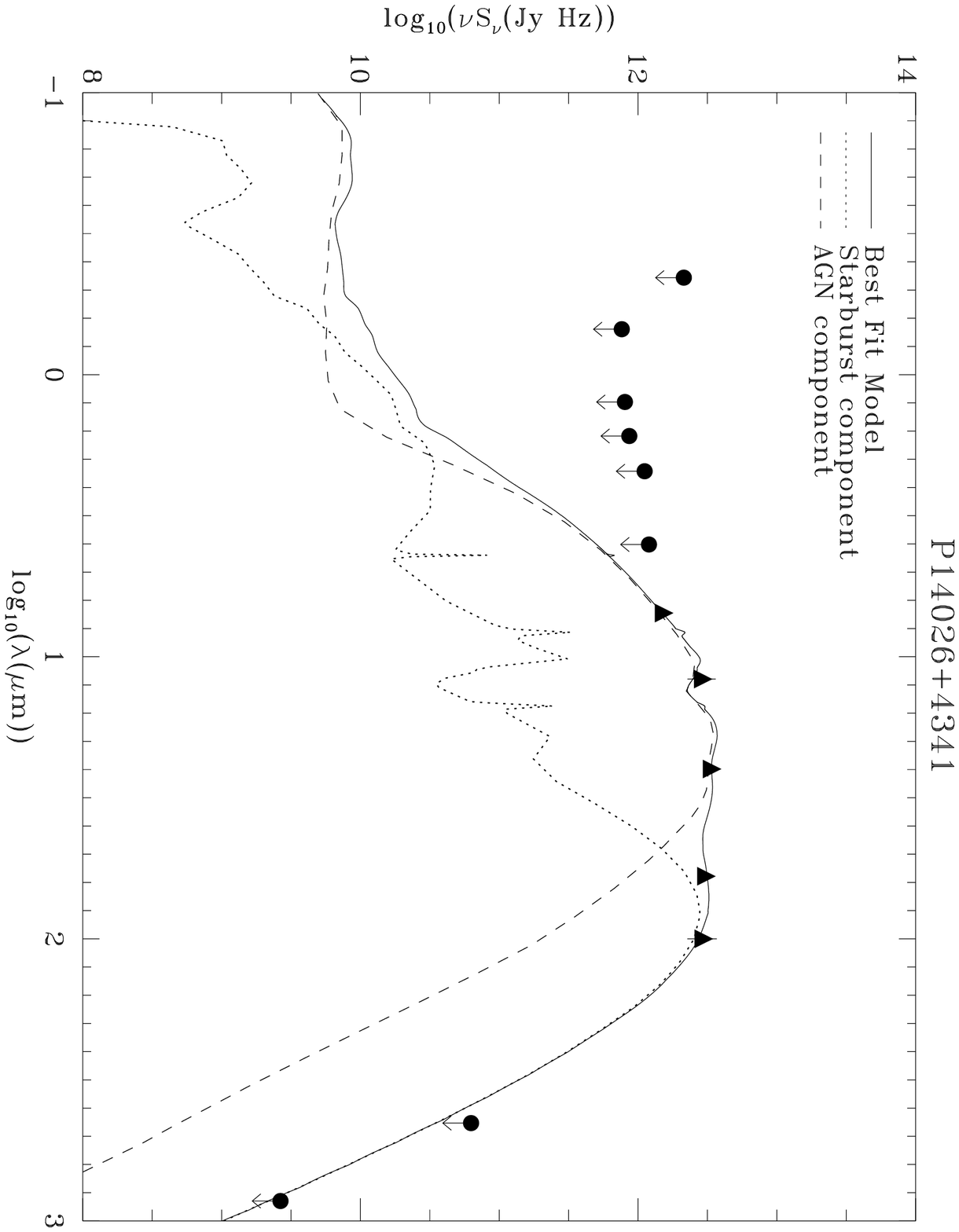,width=84mm}
\end{minipage}
\caption{Best fit combined models in the observed frame, and the AGN and Starburst components, for F00235+1024, 07380-2342, 
F12509+3122, 14026+4341, F14218+3845 \& 18216+6418
 \label{hlirg_seds1}}
\end{figure*}

\begin{figure*}
\begin{minipage}{170mm}
\epsfig{figure=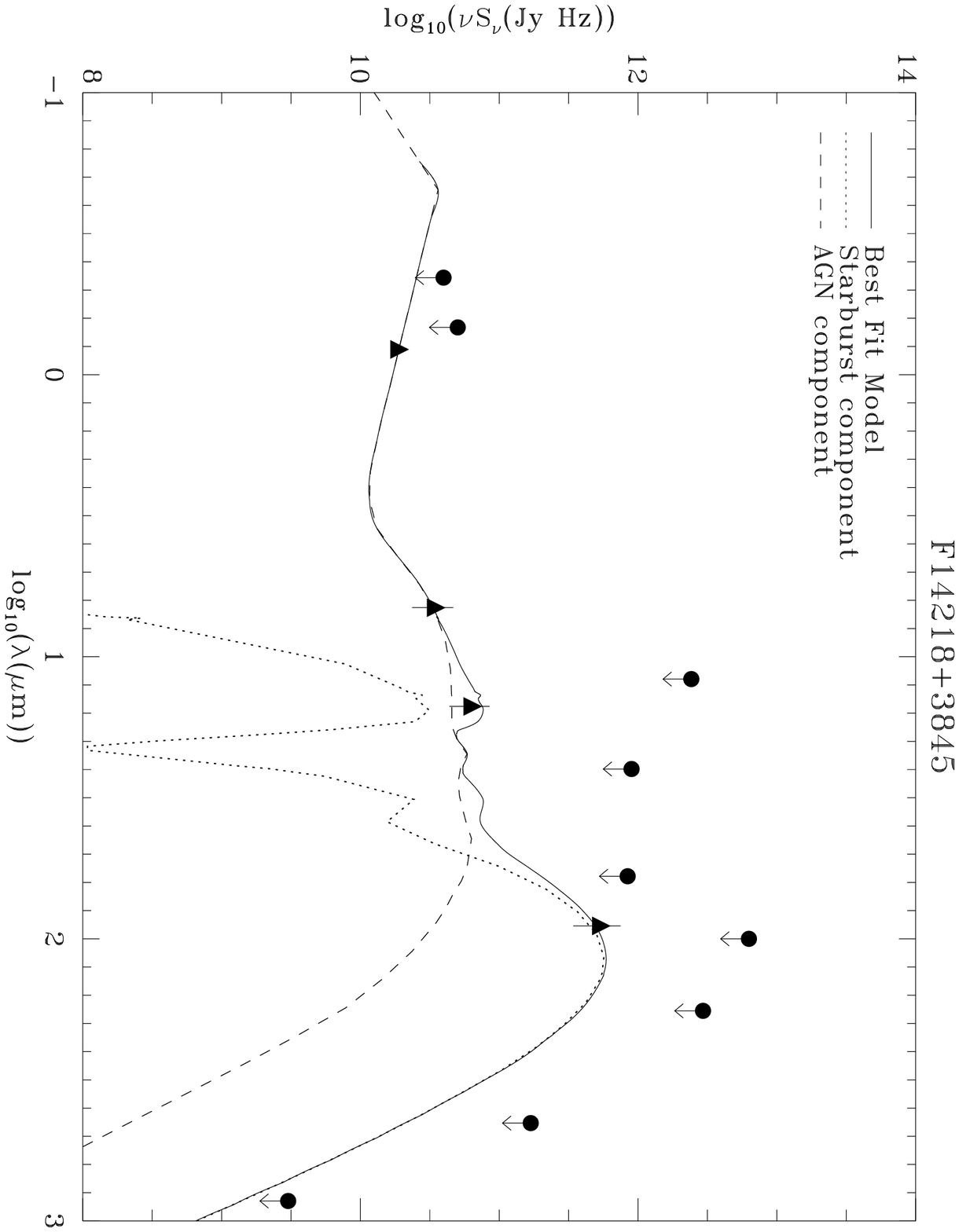,width=84mm}
\epsfig{figure=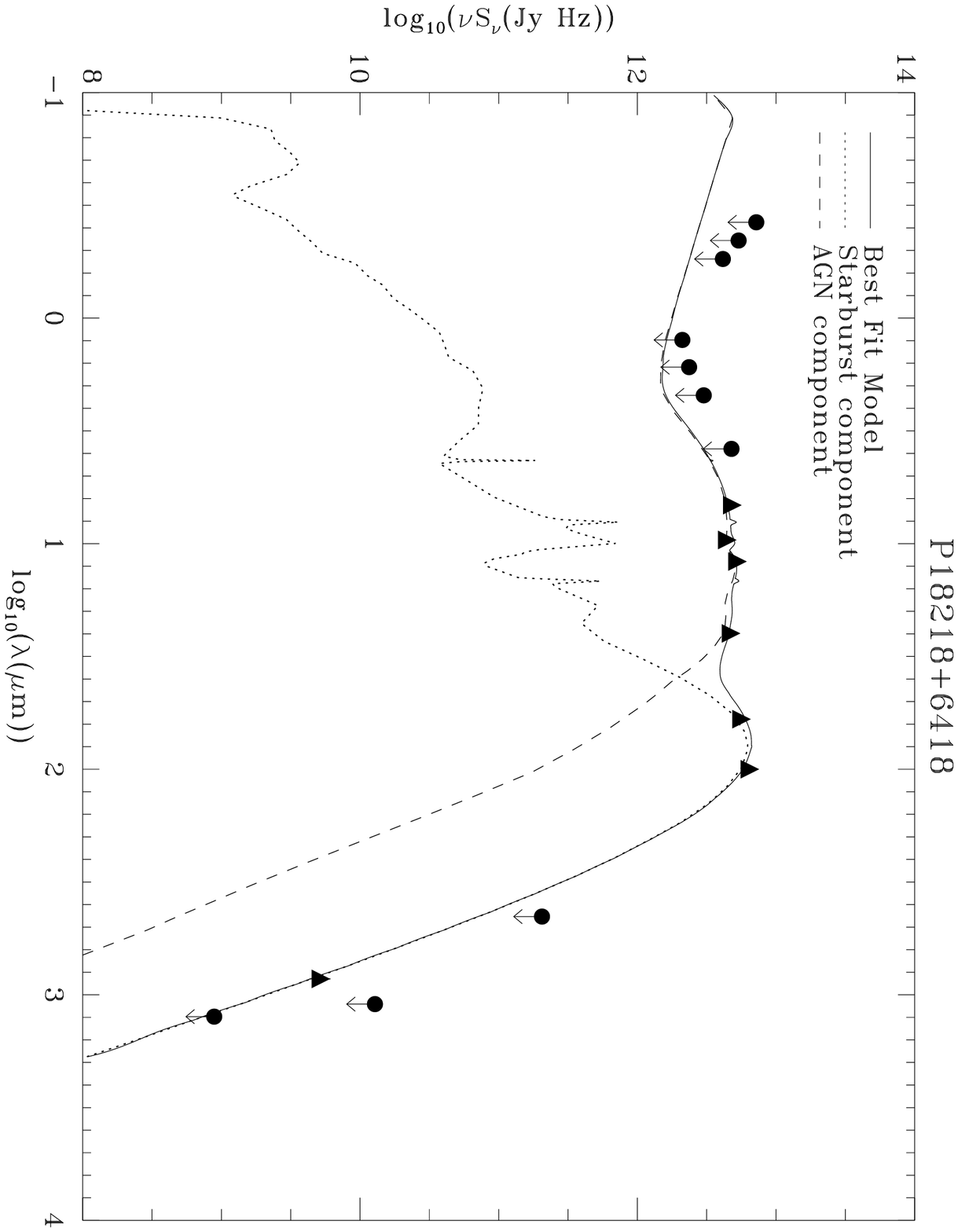,width=84mm}
\end{minipage}
\contcaption{}
\end{figure*}

\subsubsection{IRAS 07380-2342}
The best fit SED model for this object is given in Figure \ref{hlirg_seds1}. Additional {\em I} band data was taken from 
archival HST observations. This narrow line HLIRG is best fitted by a combination of starburst and AGN components where 
the AGN component dominates overall ($\sim65\%$), but where the starburst dominates at wavelengths longward of 
$45\mu$m. The IR data is best fit by a young (6.6 - 16Myr) starburst with a star formation rate of $1250M_{\odot}$yr$^{-1}$, 
and an AGN component with a line of sight angle of $<20^{\circ}$. The dust temperature and mass are 31K and 
$6.6\times10^{8}M_{\odot}$ respectively.

\subsubsection{IRAS F10026+4949}
{\it HST} imaging of this source \cite{far1} shows moderate morphological disturbance and multiple very close 
companions. The object is thus assumed to be in ongoing interactions. We compiled additional {\em I} band data for 
this object from Farrah et al. \shortcite{far1}. With only three
fluxes, there exists insufficient IR photometry to model the IR SED. The fluxes combined with the upper 
limits do however allow constraints to be drawn on both the total IR luminosity, and the starburst and AGN components. 
The best fit predicts a total IR luminosity of $\sim6.5\times10^{13}L_{\odot}$, with the AGN providing $\sim80\%$ 
of the total IR emission. The star formation rate is extremely high, at $\sim2000M_{\odot}$yr$^{-1}$.

\subsubsection{IRAS F12509+3122}
This QSO shows no signs of interaction, and has an undisturbed bright elliptical host \cite{far1}. We obtained 
additional {\em I} band data from Farrah et al. \shortcite{far1}, which is treated as a flux rather than an upper 
limit in the SED fitting as the host galaxy magnitude is known. The best fit to the IR SED predicts a total 
IR luminosity of $\sim1.82\times10^{13}L_{\odot}$, where the AGN component provides the majority ($\sim60\%$) of 
the emission over the whole SED but where the starburst dominates at wavelengths $>50\mu$m. The dust temperature 
and emissivity are 39K and 1.94 respectively, with a predicted star formation rate of $\sim1050M_{\odot}$yr$^{-1}$. 
An SED fit is plotted in Figure \ref{hlirg_seds1}. The model slightly underpredicts the $I$ band flux. This is likely 
due to treating the other optical fluxes as upper limits. As this object is a QSO the optical emission will arise 
mostly from the AGN, and indeed the host galaxy contribution in the {\it I} band is very small \cite{far1}. For
consistency with the other sources we have treated the other optical fluxes as upper limits, as trial fits 
show that fitting to the optical points as fluxes makes no difference to the derived IR luminosities. 

\subsubsection{IRAS 13279+3401}
This broad line object has only two measured fluxes, so constraints can only be drawn on the total IR luminosity, and 
the best fit AGN and starburst components. The best fit predicts a total IR luminosity of 
$\sim8.7\times10^{12}L_{\odot}$ with the AGN component providing the majority ($\sim70\%$) of the emission.

\subsubsection{IRAS 14026+4341}
This radio quiet BAL QSO was discovered by Low et al. \shortcite{low}. {\it HST} imaging shows that the host is a 
luminous elliptical, and that the host is involved in ongoing interactions \cite{hut}. Optical spectropolarimetry 
\cite{hin} shows that the polarized flux density is dependent on wavelength, indicating that the polarization is 
produced by dust scattering. 

The best fit SED for this object is presented in Figure \ref{hlirg_seds1}. For this object we obtained additional 
unpublished {\em ISO} photometry (Belinda Wilkes, priv. comm.). The best fit model predicts a total IR luminosity 
of $7.9\times10^{12}L_{\odot}$ with the AGN component providing $\sim60\%$ of the IR emission. The derived dust 
temperature and mass are 35K and $1.7\times10^{8}M_{\odot}$ respectively, with a star formation rate of 
$\sim600M_{\odot}$yr$^{-1}$.

\subsubsection{IRAS F14218+3845}
This QSO possesses a small very luminous host with no signs of ongoing interactions and an undetermined morphology 
\cite{far1}. {\em ISO} observations \cite{ver} suggest that the IR emission is best explained as being predominantly 
($74\%$) starburst in origin, but with the AGN component dominating at wavelengths $<30\mu$m.

Additional {\em ISO} photometry was compiled from Verma et al. \shortcite{ver}, and additional {\em I} band data 
from Farrah et al. \shortcite{far1}. The {\em I} band data is treated as a flux rather than an upper limit as the 
host galaxy magnitude is known. The best fit predicts a total IR 
luminosity of $1.15\times10^{13}L_{\odot}$. The starburst component dominates, providing $80\%$ of the total IR emission. 
All the model fits are consistent with a starburst age of $<57$Myr, and a dust mass and temperature of $1.07\times10^{8}M_{\odot}$ and 
45K respectively. The predicted star formation rate is extremely high, at $2000M_{\odot}$yr$^{-1}$. The best fit 
SED is presented in Figure \ref{hlirg_seds1}. The starburst component dominates at rest-frame wavelengths $>45\mu$m, with the 
AGN component dominating in the optical.

\subsubsection{IRAS F16124+3241}
This object shows no detectable polarized broad lines in the optical, suggesting that the IR emission may be powered 
by a young starburst rather than a dust shrouded AGN \cite{tra}. With only three fluxes, it is not possible to draw constraints 
on all model parameters. The best fit total IR luminosity is $\sim1.1\times10^{13}L_{\odot}$, with the starburst providing 
$\sim60\%$ of the IR emission. An AGN component is however required to explain the total IR emission over $1-1000\mu$m.

\begin{table*}
\begin{minipage}{170mm}
\caption{Hyperluminous Galaxies: Luminosities and Model Parameters \label{hlirgparams}}
\begin{tabular}{@{}lccccccccc}
\hline
Name       & $L^{Tot}_{IR}$          & $L^{Sb}_{IR}$        & $L^{AGN}_{IR}$        & Age$^{a}$ & SFR$^{b}$            & $\theta^{c}$ & $M_{d}^{d}$ & T$^{e}$ & $\beta^{f}$\\ 
           & $L_{\odot}$             & $L_{\odot}$          & $L_{\odot}$           & Myr       & $M_{\odot}$yr$^{-1}$ & \degr        & $M_{\odot}$ & kelvin  &           \\
\hline
F00235+1024& $13.04^{+0.08}_{-0.16}$ & $12.74^{+0.06}_{-0.20}$ & $12.73^{+0.11}_{-0.14}$ & --     & $1000\pm300$ & $<17.2$ & $8.29^{+0.11}_{-0.55}$ & $35\pm5$ & $1.92\pm0.1$ \\
07380-2342 & $13.34^{+0.01}_{-0.01}$ & $12.85^{+0.01}_{-0.03}$ & $13.16^{+0.01}_{-0.01}$ &$6.6-16$& $1250\pm300$ & $<20$   & $8.79^{+0.10}_{-0.21}$ & $31\pm3$ & $1.97\pm0.05$\\
F10026+4949& $13.81^{+0.03}_{-0.13}$ & $13.02^{+0.05}_{-0.14}$ & $13.73^{+0.03}_{-0.13}$ & --     & $2000\pm600$ & --      & $<8.7$                 & --       & --           \\
F12509+3122& $13.26^{+0.02}_{-0.07}$ & $12.88^{+0.09}_{-0.15}$ & $13.02^{+0.02}_{-0.07}$ & --     & $1050\pm250$ & --      & $8.20^{+0.08}_{-0.38}$ & $39\pm7$ & $1.94\pm0.07$\\
13279+3401 & $12.88^{+0.10}_{-0.05}$ & $12.39^{+0.30}_{-0.37}$ & $12.71^{+0.05}_{-0.18}$ & --     & $500\pm300$  & --      & --                     & --       & --           \\
14026+4341 & $12.90^{+0.03}_{-0.50}$ & $12.47^{+0.14}_{-0.05}$ & $12.70^{+0.03}_{-0.13}$ & $<57$  & $600\pm150$  & --      & $8.24^{+0.05}_{-0.50}$ & $35\pm5$ & $1.94\pm0.02$\\
F14218+3845& $13.06^{+0.02}_{-0.07}$ & $12.96^{+0.06}_{-0.17}$ & $12.35^{+0.10}_{-0.28}$ & $<57$  & $2000\pm500$ & $>26$   & $8.03^{+0.50}_{-0.13}$ & $45\pm10$& $1.97\pm0.02$\\
F16124+3241& $13.02^{+0.02}_{-0.14}$ & $12.79^{+0.07}_{-0.10}$ & $12.63^{+0.10}_{-0.35}$ & --     & $1100\pm300$ & --      & $8.54^{+0.1}_{-0.4}$   & $32\pm5$ & $1.94\pm0.05$\\
E J1640+41 & $12.90^{+0.01}_{-0.03}$ & $12.39^{+0.09}_{-0.04}$ & $12.73^{+0.01}_{-0.04}$ & --     & $550\pm200$  & --      & $<8.2$                 & --       & --           \\
18216+6418 & $13.14^{+0.01}_{-0.01}$ & $12.74^{+0.02}_{-0.02}$ & $12.91^{+0.01}_{-0.01}$ & $26-37$& $1100\pm200$ & $>46$   & $8.51^{+0.03}_{-0.02}$ & $31\pm4$ & $1.94\pm0.01$\\
\hline
\end{tabular}

\medskip

Luminosities are the logarithm of the 1-1000$\mu$m luminosities obtained from the best fit combined starburst/AGN model, 
in units of bolometric solar luminosities. $^{a}$Starburst age $^{b}$Star formation rate 
$^{c}$Viewing angle of the AGN dust torus $^{d}$Dust mass $^{e}$Dust temperature $^{f}$Emissivity index
\end{minipage}
\end{table*}

\subsubsection{ELAIS J1640+41}
This radio quiet QSO was discovered \cite{mor} as part of the ELAIS survey \cite{oli}. 
Unlike many other HLIRGs this object is detected in the soft X-ray, suggesting that this object may 
represent the face on analogues of narrow-line HLIRGs \cite{mor}. 

A SED fit for this object has been previously presented by Morel et al. \shortcite{mor}. Here we refit the SED 
with improved estimates for the {\it IRAS} upper limits using the SCANPI software. We derive a total IR luminosity 
of $7.9\times10^{12}L_{\odot}$ with the AGN providing $\sim70\%$ of the IR emission. The star formation rate is 
$550M_{\odot}$yr$^{-1}$. These results are in agreement with Morel  et al.

\subsubsection{IRAS 18216+6418}
This broad line object posesses some enigmatic properties. Radio observations classify it as radio 
quiet, although the richness of the environment (Abell class $\sim2$) is more typical of Radio Loud objects 
\cite{lrh}. Many other properties of the system are more typically found in radio loud objects 
The host galaxy, with a magnitude of $M_{V}=-23.2$, is a large, luminous elliptical with 
no signs of ongoing interaction. The QSO nucleus is slightly offset from the centre of the host galaxy. 
Radio mapping of this sources show a core and a pair of antipodean jets \cite{lrh,blu}, a 
radio structure more typically found in RLQs. The radio structure implies that a central supermassive 
black hole is responsible for the compact radio emission, rather than a starburst \cite{blu}. Off-nuclear 
optical spectroscopy shows a spectrum more typically found in radio loud objects \cite{fri}.

In addition to our observations of the QSO we also observed with SCUBA the other object within the IRAS error 
ellipse and found no detection, further supporting the identification of the QSO as the source of the IRAS emission.  
Additional optical photometry for this object was taken from Kolman et al. \shortcite{kol}, additional {\em ISO} 
fluxes from Clavel et al. \shortcite{cla}, and additional sub-mm data from Andreani, Franceschini \& Granato 
\shortcite{and}. The comprehensive photometry for this object allows excellent constraints on the best fit model parameters. 
The best fit total IR luminosity is $1.4\times10^{13}L_{\odot}$, with the AGN component providing $\sim60\%$ of the IR emission. 
The best fit starburst model parameters are a starburst age of between 26Myr and 37Myr, a dust mass of $3.2\times10^{8}M_{\odot}$, 
and a dust temperature of 31K. The inferred star formation rate is $1100M_{\odot}$yr$^{-1}$. 

\subsubsection{IRAS F23569-0341}
Previous {\em ISO} observations of this object \cite{ver} did not confirm the {\it IRAS} $60\mu$m detection, and did 
not detect the object in any other band. As we did not detect this source in the sub-mm, its reality remains 
to be determined. With only one IR flux, it is impossible to draw meaningful constraints on either luminosities 
or model parameters, we therefore only present a compilation of IR data for this object. Worth noting however is 
that all the model fits are consistent with a total IR luminosity of $10^{13.5}L_{\odot}$ or less.

\section{Discussion}

\subsection{Starburst \& AGN Properties}
The contributions from starburst and AGN activity to the total IR emission in the most luminous {\it IRAS} galaxies has 
been discussed by several authors. Mid-IR spectroscopy \cite{gen,rig} found that the IR emission in most ULIRGs 
($\sim80\%$) was powered by starbursts, although at least half of their samples showed evidence for both starburst and AGN 
activity. There was no detected trend for the AGN-like systems to reside in the more compact (and hence more advanced merger) 
systems. The fraction of sources powered by an AGN has been shown by several authors to increase with increasing total IR 
luminosity \cite{vei,shi,vei2}. Recent {\em ISO} observations of ULIRGs \cite{tra2} found that, at IR luminosities below 
$10^{12.4}L_{\odot}$, most ULIRGs were starburst dominated, with the starburst component contributing around $85\%$ to the 
IR emission. At IR luminosities above $10^{12.4}L_{\odot}$ the AGN contribution was much higher, contributing at least $50\%$ 
of the IR emission. 

Studies of samples of HLIRGs to look for contributions from starbursts and AGN to the total IR emission have yielded 
interesting results. Studies of small samples of HLIRGs produced conflicting results, with some authors suggesting that 
the IR emission in HLIRGs arises predominantly from a starburst \cite{fra1,fra2} with star formation rates of the order 
$10^{3}M_{\odot}$yr$^{-1}$, and other authors suggesting that HLIRGs are powered by a dusty AGN \cite{gra2,eva,yun}.
There is also evidence from spectropolarimetry for a buried AGN in several HLIRGs \cite{hin0,goo}
Previous SED modelling of HLIRGs \cite{rr2,ver} found that most required an AGN and a starburst to 
power the total IR emission. Overall, about half were AGN dominated. These results were however based on samples either 
biased toward AGN or lacking the tight constraints on starburst luminosities given by sub-mm data. 

\begin{table}
\centering{
\caption{Additional HLIRGs from the literature with calculated starburst and AGN fractions \label{addhlirgs}}
\begin{tabular}{@{}lcccccc}
\hline
Name             &  $z$   &$L^{Sb}_{IR}$ & $L^{AGN}_{IR}$ & $L^{Tot}_{IR}$ \\
                 &        & $L_{\odot}$  &   $L_{\odot}$  &    $L_{\odot}$ \\   
\hline
SMMJ02399-0136   & 2.803  & 12.89  & 12.47    & 13.03    \\
IRAS F10214+4724 & 2.286  & 13.10  & 13.27    & 13.49    \\
PG1148+549       & 0.969  & 13.27  & 13.51    & 13.71    \\
IRAS F12514+1027 & 0.30   & 12.67  & 12.63    & 12.95    \\
H1413+117        & 2.546  & 12.71  & 13.38    & 13.46    \\
IRAS F14481+4454 & 0.66   & 12.94  & 13.03    & 13.29    \\
IRAS F15307+3252 & 0.93   & 12.96  & 13.28    & 13.45    \\
\hline
\end{tabular}

\medskip

Luminosities are given for $H_{0}=65$ km s$^{-1}$ Mpc$^{-1}$, $\Omega_{0}=1.0$. The luminosities of 
SMMJ02399-0136, F10214+4724 and H1413+117 have been corrected for the effects of gravitational lensing. Taken 
from Rowan-Robinson (2000) \& Verma et al. (2002).
}
\end{table}

The total IR luminosities of the 10 reliably detected objects in our sample, together with 
the starburst and AGN fractions, are presented in Table \ref{hlirgparams}. In all ten objects both a starburst 
and an AGN are required to explain the total IR emission; in all cases a `pure' starburst or AGN model is unable 
to explain the emission over $1-1000\mu$m. The mean starburst fraction in our sample is $\sim35\%$. In all ten objects 
both the AGN and starburst components supply at least $20\%$ of the total IR luminosity, any of the objects 
would be classified as at least a ULIRG with only the starburst or AGN component present. Both starburst and AGN 
activity in HLIRGs are therefore central in interpreting their properties.  

\begin{figure*}
\rotatebox{0}{
\centering{
\scalebox{0.5}{
\includegraphics*[18,144][592,718]{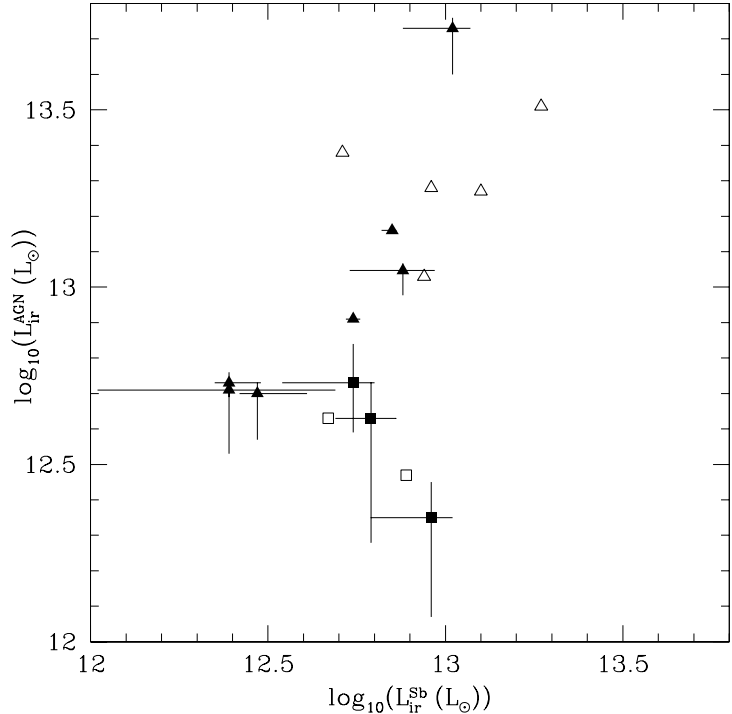}
}}}
\caption
{
AGN luminosity in the IR plotted against starburst luminosity in the IR. Triangles are AGN dominated 
systems, squares are Starburst dominated systems. Filled shapes are objects from the sample in this paper, outline shapes are 
taken from Table \ref{addhlirgs}.
\label{sbvsagn}
}
\end{figure*}

Figure \ref{sbvsagn} shows a plot of bolometric AGN luminosity plotted against bolometric starburst luminosity for our sample. 
Also included in this figure are those HLIRGs from Rowan-Robinson (2000) and Verma et al. (2002) that have constrained
starburst and AGN fractions. These additional HLIRGs are listed in Table \ref{addhlirgs}. It should be noted that only two 
of these objects (SMMJ02399-0136 \& F10214+4724) are selected in a manner unbiased toward AGN. Three of the objects in Table 
\ref{addhlirgs} are gravitationally lensed (SMMJ02399-0136, F10214+4724 \& H1413+117), and the effects of lensing must be 
corrected for to allow comparisons with unlensed objects. For SMMJ02399-0136 we have used the lensed starburst and AGN 
luminosities given by Rowan-Robinson \shortcite{rr2} and applied a lensing correction of 2.5 \cite{ivi}. For F10214+4724 we 
have taken the unlensed starburst and AGN luminosities from Green \& Rowan-Robinson \shortcite{gr2}. For H1413+117 (the 'Cloverleaf') 
we have taken the lensed starburst and AGN luminosities given by Rowan-Robinson \shortcite{rr2} and applied a lensing correction 
of a factor of 10 \cite{yun0}.

It can be seen from Figure \ref{sbvsagn} that there is a wide spread in fractional power sources across the sample, 
ranging from $80\%$ starburst dominated to $80\%$ AGN dominated. This is consistent with a similar plot given in Rowan-Robinson 
\shortcite{rr2} and confirms that a wide spread in fractional starburst luminosities exists over at least 5 orders of magnitude 
in total IR luminosity in the IR galaxy population. The measured star formation rates are all greater than $500M_{\odot}$yr$^{-1}$, 
and in seven cases exceed $1000M_{\odot}$yr$^{-1}$. For the six objects in the sample where the SED shape is constrained, the 
IR emission longward of $50\mu$m in the rest-frame is in all cases dominated by star formation. 

Further deductions can be made from Figure \ref{sbvsagn}. With only 5 systems in which the starburst contributes more to the 
total IR luminosity than the AGN it is impossible to draw conclusions on any correlation between starburst and AGN luminosity. Amongst the 
12 AGN dominated systems however there is a strong correlation between the starburst and AGN luminosities. Further comparison with figure 
19 from Rowan-Robinson \shortcite{rr2} shows that this correlation extends over 5 orders of magnitude in total IR 
luminosity. Considering only the AGN dominated HLIRGs in Figure \ref{sbvsagn}, the Spearman rank correlation coefficient between 
the AGN luminosity and the starburst luminosity is 0.95. When the additional AGN dominated systems from Table \ref{addhlirgs} 
are included the correlation coefficient becomes 0.98. 

This is an interesting result for understanding AGN and starburst activity in {\it IRAS} galaxies at all luminosities, as it implies 
that the luminosities of the AGN and starburst components are influenced by common physical factors. On {\em prima facie} 
grounds this may seem counterintuitive. Recent studies of starburst and QSO triggering in mergers between gas rich galaxies 
\cite{bar3,ta,bek} suggest a complex pattern of starburst activity. Depending on the impact parameters of the merger, the number 
of merger progenitors and their morphologies, there may be multiple starburst events triggered throughout the lifetime of the 
merger. AGN activity conversely, is likely to follow a simpler pattern. If a $>10^{8}M_{\odot}$ black hole is not located in 
either of the merger progenitors then any central AGN 
will not reach QSO level luminosities until at least $10^{8}$ years after the merger has commenced \cite{ta}, leading 
to a potential lag between any {\em single} starburst and AGN event. It may thus be expected that there should be at best  
a weak correlation between starburst or AGN luminosities against total IR luminosities in interaction driven {\it IRAS} galaxies. 
Our results however suggest that there is an underlying factor linking the luminosities of the starburst and AGN components. 
As the physical mechanisms behind star formation and AGN activity are different, the most likely candidate is the 
available quantity of fuel in the circumnuclear regions of the system. The same reasoning applies if we consider HLIRGs 
as young galaxies in the process of formation.

\begin{figure*}
\rotatebox{0}{
\centering{
\scalebox{0.5}{
\includegraphics*[18,144][592,718]{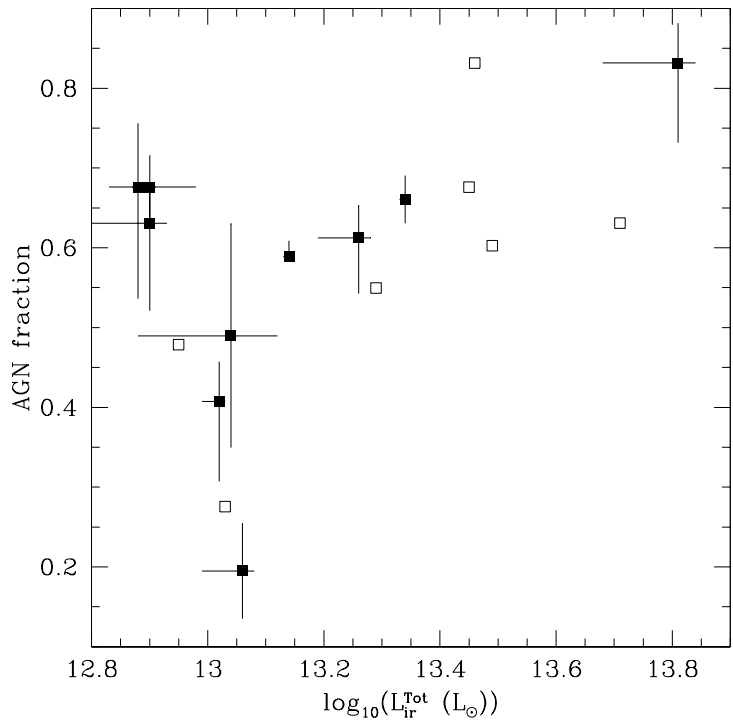}
}}}
\caption
{Fraction of the IR luminosity due to AGN activity against the total IR luminosity. Filled squares are objects from this 
sample, outline squares are the additional HLIRGs from Table \ref{addhlirgs}. 
\label{agnfrac}
}
\end{figure*}

The fraction of the IR luminosity due to AGN activity plotted against the total IR luminosity is given in Figure \ref{agnfrac}. 
Also plotted are the additional HLIRGs from Table \ref{addhlirgs}. It can be seen that there is no 
clear trend of increasing AGN fraction with increasing IR luminosity amongst the HLIRG population, irrespective of whether 
biased or unbiased samples are selected. We therefore find that the trend observed in the ULIRG population of increasing AGN 
fraction with increasing IR luminosity peaks in the HLIRG population, and that the AGN fraction does not continue 
to increase at IR luminosities greater than $10^{13}L_{\odot}$. 

This study has demonstrated that, to accurately gauge both the total IR luminosity in the most luminous {\it IRAS} sources, and the contributions 
from starburst and AGN components, it is necessary to have fluxes and upper limits from the near-IR to the sub-mm. Two
objects in our sample (F00235+1024, F14218+3845) have different IR luminosities, and starburst and AGN fractions, than 
those derived by previous authors \cite{rr2,ver} without sub-mm fluxes or upper limits.

\subsection{Evolutionary Models for HLIRGs}
A prevalent picture for the evolution of the most luminous {\it IRAS} galaxies is the Sanders et al. \shortcite{san2} 
picture. Under this picture (hereafter the S88 picture), ULIRGs are the dust shrouded 
precursors to optical QSOs. Interactions and mergers between gas rich spirals transport gas to the 
central regions of the galaxies. This large central gas concentration triggers starburst activity, and in the 
latter stages of the merger commences the fuelling of a central supermassive black hole. In the last stages the dust 
screen shrouding the black hole is blown away and the ULIRG evolves into an optical QSO. A more recent evolutionary 
scenario for ULIRGs \cite{far2} is that ULIRGs are not a simple precursor stage to optically selected 
QSOs, but instead are a diverse population whose evolution is driven by the progenitor morphologies and 
the local environment. HLIRGs may simply be the high luminosity 'tail' of the ULIRG population. Also worth noting is 
that there is plausible evidence \cite{far1} that HLIRGs may lie in very diverse environments, from isolated sources 
to Abell class $\geq$2 clusters. Such a diversity of environment would give rise to a more diverse range 
of evolutionary paths. 

A recent model for the evolution of HLIRGs in particular is that of Rowan-Robinson \shortcite{rr2}. According to this 
model, HLIRGs are ideal candidates for being `primeval', or very young galaxies, as opposed to mergers between gas rich 
galaxies. Evidence in support of this is that HLIRGs all have very high star 
formation rates and a higher gas fraction than typically found in spiral galaxies. The data in this paper, coupled with 
previous results, allows us to examine whether HLIRGs are very luminous galaxy mergers, or young active galaxies. 

The amounts of gas and dust in active galaxies have been examined by several authors and can be a useful diagnostic. In the 
local Universe, observations of nearby radio galaxies in the sub-mm \cite{kna} found some objects to have comparable dust 
masses ($\sim10^{7}M_{\odot}$) to those found in spiral galaxies, but also found other objects with dust masses as low as 
$\sim10^{4}M_{\odot}$. At higher redshifts, studies of samples of radio galaxies and radio quiet quasars \cite{hdr,mcm,and} 
derive dust masses in the range $10^{7} < M_{dust}(M_{\odot}) < 10^{9}$, with no appreciable dependence on redshift out to $z\sim5$.
These results imply an early accelerated phase of obscured star formation in high redshift primeval galaxies \cite{fran,maz}.
Gas and dust masses in HLIRGs have also been the subject of several previous studies, with derived dust masses lying in the 
range $10^{8} < M_{dust}(M_{\odot}) < 10^{9}$ \cite{bac,dow,fra1,fra2,ver}.

The dust masses derived for the sources in our sample are presented in Table \ref{hlirgparams}. 
Assuming that the gas to dust ratio in HLIRGs is similar to that in ULIRGs, adopting a gas to dust ratio of 540 \cite{sand} we 
derive gas masses in the range $10^{10.5} < M_{gas}(M_{\odot}) < 10^{11.5}$. These gas and dust masses are comparable to those 
found by previous authors in other HLIRGs, and in other classes of active galaxy. A young, active galaxy may be expected to have 
both a very high rate of star formation ($\geq500M_{\odot}$yr$^{-1}$) in order to form the 
bulk of its' stars in $\sim1$Gyr, and a large gas reservoir ($M_{gas}\sim10^{11}M_{\odot}$) to maintain this rate of star 
formation. Both these criteria are satisfied by all the objects in our sample. In addition, several of the host galaxies in our 
sample show both an elliptical morphology and no signatures of ongoing interaction found in mergers between equal mass spiral galaxies 
\cite{far1}. They are therefore excellent candidates for young, active galaxies. 

Some of our sample however have optical morphologies that do not fit with this scheme. {\em HST} imaging \cite{far1} has shown that some HLIRGs, such 
as F00235+1024 and F10026+4949 lie in interacting systems with similar morphologies to local ULIRGs. The most natural 
interpretation of these sources is that they are the high luminosity 'tail' of the ULIRG population. It is also 
conceivable that the starburst and AGN activity in the dynamically relaxed systems in our sample are also triggered by galaxy 
interactions. The dynamical relaxation time of a merger between galaxies of comparable size is of the order $\sim10^{9}$ 
years \cite{bar3,dub}, whereas the lifetime of a QSO \cite{ree,mar} and a starburst \cite{gen,thor} is thought to be no 
more than $10^{8.5}$ years. It is therefore just possible that a merger driven coeval starburst and AGN could be observed 
in a dynamically relaxed host, if the starburst and AGN activity were not triggered at or near the start of the merger. More 
likely though is that the objects in question are very young galaxies going through their maximal star forming period. We 
therefore conclude that the HLIRG population is comprised of both the most 
luminous end of the merger driven ULIRG population, and of a population of young active galaxies going through their 
maximal star formation periods.

\subsection{The Cosmic X-ray and Sub-mm Backgrounds}

The first three HLIRGs to be discovered were all shown to harbour obscured AGN \cite{hin0}. Just as Seyfert 2 galaxies 
have been suggested to be Seyfert 1 galaxies viewed along a line of sight where the central regions are obscured by dusty 
tori, HLIRGs have been proposed as `misdirected' QSOs, or QSO-2's \cite{hin1}. A similar picture has also been proposed 
for radio galaxies \cite{ali}. Under this picture, HLIRGs would be classified as Hyperluminous quasars if viewed along a 
polar line of sight, and as Hyperluminous Galaxies if viewed under any other orientation. 

We have shown that all the sources in our sample require extremely luminous AGN to explain the total IR 
emission. Two objects, IRAS 18216+6418 and ELAIS J1640+41, were previously known to harbour AGN, since they are both luminous in 
the hard X-ray \cite{prm,mor}. On the other hand, deep X-ray observations of 5 other HLIRGs \cite{fab1,wi}, including one (F00235+1024) from 
our sample, did not detect any of the objects, from which it was concluded that these objects were either atypically weak in the 
X-ray or that the central regions were obscured by columns $N_{H}>10^{23}$cm$^{-2}$. From the dust masses and AGN IR luminosities  
presented in this paper the latter conclusion is preferred. As the sample selection for our study is independent of AGN content, it 
seems likely therefore that many HLIRGs as a class contain a luminous AGN that is heavily obscured and can only be detected in the IR. 

The presence of highly obscured AGN in a large fraction of the most luminous {\it IRAS} galaxies has interesting implications for interpreting 
observations of the cosmic X-ray and IR backgrounds. Studies have shown that only $\sim10\%$ of high-$z$ bright sub-mm sources show 
signs of AGN activity at other wavelengths \cite{barg,alma}. Recent surveys have shown that the hard X-ray background is due almost 
entirely to AGN \cite{mus}, and there exist models for the X-ray background in which low luminosity AGN produce a hard X-ray 
spectrum via photoelectric absorption from an obscuring starburst \cite{fab2,gun}. Estimates of the AGN contribution to deep sub-mm surveys and the 
IR background \cite{alb} assume that the fractions of obscured and unobscured AGN must fit with the observed cosmic 
X-ray background, and from this estimate that between $10\%$ and $20\%$ of the $850\mu$m SCUBA sources at 1mJy, and 
a similar fraction of the IR/sub-mm background, are due to AGN. 

From our results however it is likely that, at least for the rest-frame most IR luminous sub-mm sources, this picture is 
not true, and that there exists a population of heavily obscured AGN in sources with $L_{ir}>10^{12.5}L_{\odot}$. These AGN will 
contribute negligibly to the cosmic X-ray background. From the SEDs presented in this paper it is apparent 
that these AGN may supply a significant fraction, possibly up to $50\%$, of the contribution to the cosmic background radiation 
at $\lambda_{obs}\geq300\mu$m from the most luminous sub-mm sources. Worth noting however is that these SEDs also predict that 
the $850\mu$m background from the most luminous sub-mm sources will in all cases be due to star formation, with a negligible 
AGN contribution.

\begin{table}
\centering{
\caption{Predicted HLIRG SCUBA fluxes extrapolated to higher redshifts \label{hlirgpredz}}
\begin{tabular}{@{}l|cccccc}
\hline
Name        & \multicolumn{2}{c}{z=2.0} & \multicolumn{2}{c}{z=3.0} \\
            & $S_{850}$ & $S_{450}$     & $S_{850}$ & $S_{450}$     \\
\hline
F00235+1024 & 5.3       & 28.0          &  6.7      &  21.2         \\    
07380-2342  & 13.1      & 59.4          &  15.2     &  47.5         \\    
F12509+3122 & 3.0       & 15.3          &  3.7      &  14.9         \\   
14026+4341  & 2.9       & 15.0          &  3.5      &  13.4         \\     
F14218+3845 & 5.6       & 37.0          &  7.2      &  38.5         \\         
18216+6418  & 8.0       & 40.5          &  9.7      &  34.7         \\     
\hline
\end{tabular}

\medskip

All fluxes are given in mJy. Predicted fluxes are given only for those 6 sources with constrained SED shapes.
}
\end{table}

\subsection{Implications for Sub-mm Surveys}

\subsubsection{Coeval starburst and AGN activity}

There exists substantial observational evidence that, at both low and high redshifts, the main episode(s) of star formation 
in galaxies are linked with QSO activity. At low redshift there is an observed correlation between black hole mass and bulge mass 
\cite{mag,kog} with a linear scaling between the two quantities \cite{mcl}.
At high redshift studies of emission and absorption lines in high-$z$ QSOs typically show 
that they reside in metal rich (at least $1Z_{\odot}$) environments, and that the more luminous objects may be more metal rich
\cite{hff}. Sub-mm continuum observations of high-$z$ QSOs generally show the presence of large masses of cold dust 
\cite{omo,ben}. This emission from cold dust is generally ascribed to star formation, although QSO activity is also 
a possibility \cite{car}. The high metallicities in these systems is often seen to indicate that the optical QSO occurs 
after the bulk of star formation has occurred in the system. A recent study of QSOs and their hosts \cite{gra3} proposes that, 
in an optical QSO, the star formation is effectively at an end as the QSO power 
and feedback from supernovae combine to blow out any nuclear starburst \cite{sil}. 
An optical QSO activity and high star formation rates are thus predicted not to be coeval. A similar study \cite{arc} predicts 
that, unless a massive `seed' black hole is already present, luminous AGN activity cannot arise until $0.5$Gyr after the 
first stars form. An independent estimate for the formation of a $>10^{8}M_{\odot}$ black hole in non-nucleated sources \cite{ta} 
gives a timescale of 1Gyr. Therefore, this predicts that there is a time lag between the peak in sub-mm luminosity (due to the 
starburst) and mid-IR luminosity (due to the AGN) of $\sim10^{8}$ years. AGN activity is predicted to only occur in the host 
once $\sim90\%$ of the stars in the host have formed and the sub-mm luminosity is only $\sim25\%$ of its peak value. The predictions 
from Archibald et al. \shortcite{arc} are that AGN in luminous high-$z$ sub-mm sources are $\sim1000$ times less luminous 
than the most luminous AGN found in other systems. 

The results from our study however suggest that, at least for the most luminous IR galaxies, coeval AGN activity and massive 
starburst activity exist in many sources, including optical QSOs. Three objects in our sample (F12509+3122, F14218+3845 \& ELAIS 
J1640+41) are optical QSOs containing an extremely luminous AGN together with high rates ($>500M_{\odot}$yr$^{-1}$) of star 
formation. The computed star formation rates are so high that it is unphysical to argue that they are not at or near the peak 
rates of star formation in these systems. In addition the strong correlation between the starburst and AGN 
luminosities in our sample, implying a common factor governing the luminosities, suggests that the starburst and AGN phases 
cannot be widely separated in time in HLIRGs.

\subsubsection{The nature of high-z sub-mm sources}

Two SCUBA galaxies have been previously shown to be HLIRGs \cite{ivi1}. A sample of HLIRGs with selection unbiased toward AGN 
allows us to examine what fraction of HLIRGs as a class would be detected by SCUBA surveys if they lay at higher redshifts. Analysis can 
only be done for those six sources in the sample (Figure \ref{hlirg_seds1}) with constrained SED shapes. The fluxes predicted to be 
observed by SCUBA for these sources if the sources lay at $z=2$ and $z=3$ are given in Table \ref{hlirgpredz}. At these redshifts 
all the sources would have observed frame $850\mu$m and $450\mu$m fluxes comparable to the range of $850\mu$m and $450\mu$m fluxes 
for sources found in recent sub-mm surveys \cite{barg0,eal,sco,fox}. Thus it seems plausible that at least some fraction of 
the high-$z$ sources found in sub-mm surveys are similar in nature to the galaxy populations found in the HLIRG population, namely a 
mixture of ULIRG-like galaxy-galaxy mergers and young active galaxies. It is thus plausible from our results that many of the sources 
discovered in sub-mm surveys harbour both a dusty starburst and a heavily obscured AGN which contribute comparably to the total 
rest-frame IR emission. Our study has also shown that, at rest-frame wavelengths $\geq50\mu$m the emission from our sample is starburst 
dominated. It is therefore a safe assumption that high-$z$ surveys sampling rest-frame wavelengths longward of $50\mu$m are for 
all sources sampling the starburst dominated region of the SED.

It is, however, important to note that X-ray observations of HLIRGs and X-ray surveys of high-$z$ sub-mm sources imply that the two 
populations are not identical, and that there has been evolution between the two epochs. Of the seven HLIRGs at $z<1$ that 
have deep X-ray observations two (E J1640+41 \& 18216+6418) are detected, corresponding to a detection fraction of $28\%$. Conversely, 
deep Chandra observations of the SCUBA 8mJy survey regions \cite{alma} detect only 2/36 sub-mm sources (assuming a detection threshold 
of $3.5\sigma$ for the sub-mm sources), or $\sim6\%$. When combined with our results this gives rise to two possibilities. The first 
is that the fraction of sources that contain luminous AGN is comparable at $z\sim1$ and $z\sim3$, with a greater mean obscuration 
level at $z\sim3$. The second possibility is that the fraction of sources that contain luminous AGN has declined from $z\sim3$ 
to $z\sim1$. Further deep observations of HLIRGs and high-$z$ sub-mm sources in the X-ray and in the IR will be required to differentiate 
between these two possibilities. 

\section{Summary}
We have presented sub-mm photometry for 11 Hyperluminous Infrared Galaxies selected purely on the 
basis of their IR emission, and used radiative transfer codes for starbursts and AGN in conjunction with IR photometry 
from the literature to examine the power source behind the IR emission from 10 of the objects. Our 
conclusions are:

\noindent (1) In all of the sources both a starburst and AGN component are required to explain the total IR emission. The mean 
starburst fraction is $35\%$, with a wide range spanning $80\%$ starburst dominated to $80\%$ AGN dominated. In all cases the 
starburst dominates at rest-frame wavelengths longwards of $50\mu$m, with associated star formation rates of at least 
$500M_{\odot}$yr$^{-1}$. 

\noindent (2) The trend of increasing AGN fraction with increasing IR luminosity observed in the ULIRG population is observed 
to peak in the HLIRG population. The correlation between the starburst and AGN IR luminosities in the AGN dominated 
systems implies that the luminosities of both starbursts and AGN in HLIRGs are governed by common physical factors, the most 
plausible of which is the available quantities of gas and dust in the system

\noindent (3) Comparisons between the fractional AGN and starburst luminosities, derived dust masses, and previously published 
HST imaging suggests that the HLIRG population is comprised both of mergers between gas rich galaxies, as found in the 
ULIRG population, and of young active galaxies going through their maximal star formation periods whilst harbouring an AGN. 

\noindent (4) The presence of a luminous AGN in all of the objects in our sample, coupled with the 
non-detection in the X-ray of most HLIRGs, implies that many galaxies with $L_{ir}>10^{12.5}L_{\odot}$ 
harbour a heavily obscured AGN. Although the contribution from these bright sub-mm galaxies to the total cosmic sub-mm 
background is small, the contribution from heavily obscured AGN in the brightest sub-mm sources to the cosmic background 
radiation at $\lambda_{obs}\geq300\mu$m may therefore be higher than estimates based on constraints from the cosmic 
X-ray background. The $850\mu$m background from the most luminous sub-mm sources will however in all cases be dominated by 
star formation. 

\noindent (5) The detection of coeval AGN and starburst activity in our sources coupled with the strong correlation between 
starburst and AGN luminosity implies that, for the most luminous sub-mm sources, star formation does not cease once the 
source harbours a luminous AGN. We infer that starburst and AGN activity, and the peak starburst and AGN luminosities, can be 
coeval in active galaxies generally. 

\noindent (6) When extrapolated to high-$z$ our sources have comparable observed frame sub-mm fluxes to objects found in 
sub-mm surveys. At least some high-$z$ sub-mm survey sources are therefore likely to be comprised of similar galaxy 
populations to those found in the HLIRG population, namely galaxy-galaxy mergers and young active galaxies. It is also 
plausible from these results that high-$z$ sub-mm sources harbour heavily obscured AGN. It should be noted that $\sim28\%$ 
of HLIRGs at $z\stackrel{<}{_\sim}1$ are detected in the X-ray, whereas only $\sim6\%$ of high-$z$ sub-mm sources are detected 
in the X-ray. This implies some evolution between the two epochs. Either the mean AGN obscuration level is 
greater at $z\sim3$ than at $z\sim1$, or the fraction of IR-luminous sources at $z\sim3$ that contain AGN is smaller than that 
at $z\sim1$.

\section{Acknowledgments}
We would like to thank Elese Archibald for invaluable help at the JCMT,  Matthew Fox and Susie 
Scott for help with the data reduction and analysis, Jose Afonso for helpful discussion, Kate Isaak 
for performing the January 2001 observations, Belinda Wilkes for supplying unpublished IR photometry 
for IRAS 14026+4341, and the anonymous referee for very helpful comments. 

The JCMT is operated by the Joint Astronomy Centre on behalf of the UK Particle Physics and Astronomy 
Research Council, The Netherlands Organization for Scientific Research and the Canadian National Research 
Council. The work presented has made use of the NASA/IPAC 
Extragalactic Database (NED), which is operated by the Jet Propulsion Laboratory under contract with 
NASA, and the Digitized Sky Surveys, which were produced at the Space Telescope Science Institute under 
U.S. Government grant NAG W-2166. The images of these surveys are based on photographic data obtained 
using the Oschin Schmidt Telescope on Palomar Mountain and the UK Schmidt Telescope. This research has made 
use of the NASA/IPAC Infrared Science Archive, which is operated by the Jet Propulsion Laboratory, California 
Institute of Technology, under contract with the National Aeronautics and Space Administration. D.G.F  
acknowledges the award of tuition fees and maintenance grant provided by the Particle Physics and 
Astronomy Research Council. This work was in part supported by PPARC (grant number GR/K98728).

\bsp 

\label{lastpage}

\end{document}